\documentclass[aip, amsmath,amssymb, reprint]{revtex4-2}
\usepackage[T1]{fontenc}
\usepackage{newpxtext}                     % Palatino body text
\usepackage{eulervm} % Euler math font, clean and upright
\usepackage{mathtools}
\DeclarePairedDelimiterX\braket[2]{\langle}{\rangle}{#1 \delimsize\vert #2}
\DeclarePairedDelimiterX\matrixel[3]{\langle}{\rangle}{#1 \delimsize\vert #2 \delimsize\vert #3}
\DeclarePairedDelimiterX\comm[2]{[}{]}{#1, #2}
\DeclarePairedDelimiterX\acomm[2]{\{}{\}}{#1, #2}

\usepackage{bm}                            % Bold math variables
\usepackage{siunitx}                       % Modern usage: \qty{\sim 200}{\kelvin}
\usepackage{derivative}
\usepackage{mleftright}
\mleftright

\usepackage{graphicx}
\usepackage{dcolumn}                       % Align table columns on decimal point
\newcolumntype{d}[1]{D{.}{\cdot}{#1}}
\usepackage{tikz}
\usetikzlibrary{decorations.markings}
\usetikzlibrary{arrows.meta}               % for Slater-diagram electron arrows
\usepackage{xcolor,soul}                   % Text highlighting
\usepackage{hyperref}
\begin{document}
%\title{Towards an understanding of charge-transfer state formation in Y6 dimers from DFTB Surface Hopping simulations}
\title{Charge-transfer state formation in Y6 (BTP-4F) dimers from TD-DFTB surface hopping}

\author{Isabel Creed}
\affiliation{Department of Chemistry, Imperial College London, Exhibition Road, London  SW7 2AZ, UK}
\author{Gonzalo Díaz Mirón}
\affiliation{ The Abdus Salam International Centre for Theoretical Physics, Trieste}
\author{Emily Wentworth}
\affiliation{Department of Chemistry, Imperial College London, Exhibition Road, London  SW7 2AZ, UK}

\author{Wenbo Sun}
\affiliation{Institute for Theoretical Physics and Bremen Center for Computational Materials Science,University of
Bremen}
\author{Carlos R. Lien-Medrano}
\affiliation{Institute for Theoretical Physics and Bremen Center for Computational Materials Science,University of
Bremen}

\author{Jarvist Moore Frost}
\affiliation{Department of Chemistry, Imperial College London, Exhibition Road, London  SW7 2AZ, UK}
\affiliation{Department of Physics, Imperial College London, Exhibition Road, London  SW7 2AZ, UK}
\email[Electronic mail:]{jarvist.frost@imperial.ac.uk}

\date{\today} % Leave empty to omit a date

\begin{abstract}
Neat films of the non-fullerene acceptor Y6 (BTP-4F) separate charge with remarkably low energetic loss, a behaviour widely attributed to intermolecular charge-transfer (CT) states whose ultrafast formation depends sensitively on local molecular packing.
To resolve this packing dependence with atomistic detail, we perform hybrid time-dependent density functional tight binding (TD-DFTB) fewest-switches surface hopping on the six Y6 crystal contact-pair dimers during the first 100~fs following photoexcitation.
Analysing the evolving transition density matrices across the donor and acceptor moieties shows that every dimer rapidly relaxes toward $S_1$, but that the specific dimer packing dictates the nature of the generated charge-transfer state. 
Donor--acceptor overlap structures directly generate intermolecular CT states, whereas acceptor--acceptor overlapping structures generate states trapped in a mix of intra- and intermolecular excitation. 
Capturing this non-equilibrium partitioning requires explicit atomistic dynamics beyond static calculations or effective models, though we find that analysing a thermal ensemble of structures (without nonadiabatic dynamics) provides some of the key information. 
We suggest the different location of these states as providing a microscopic rationale for morphology-controlled charge separation in Y6 solids.
\end{abstract}


\maketitle

\section{Introduction}

The development of non-fullerene (electron) acceptors (NFAs) has led to a significant increase in organic photovoltaic (OPV) power-conversion efficiency. 
The archetypical modern NFA is Y6 (BTP-4F)\cite{Yang2021}, which has a banana-shaped structure, with a flat A-DA'D-A structure, with $\pi$--$\pi$ stacking of the fused cores required for high inter-molecular mobility.
Y6 exhibits good processing behaviour with a wide variety of co-deposited materials, has strong near-infrared optical absorption of its own, and seems to possess good charge-carrier transport properties\cite{yuan2019single}. 
Overall, Y6 has become the first choice electron-acceptor in a bulk heterojunction organic photovoltaic.

Neat Y6 films demonstrate unusual photophysics. 
The solvatochromic shift of Y6 in the solid-state is very large \cite{Regnier2023}. 
There is evidence that free charges may be directly generated without an applied bias\cite{price2022free}; certainly the energetic cost of exciton dissociation into free charges is small\cite{Zhu2022,Hart2026}. 
As a simple bilayer with rubrene, Y6 upconverts infrared light into visible (via triplet-triplet annihilation; two photons in for one photon out) with the highest efficiency for a solid-state device\cite{izawa2021efficient}.

The concept of intermolecular charge transfer (CT) states\cite{si2023photo, price2022free,izawa2021efficient, lan2024correction} is often invoked to explain the unusual photophysics of Y6, particularly how it seems to enable fast and low-energy charge separation, and high upconversion yield. 
As these charge-transfer states are intermolecular, a structural model of packing is needed, and must, at the very least, consider dimers as a model for the solid state. 
Early studies required this to come from molecular dynamic simulations\cite{Zhang2020}, but more recent works\cite{giannini2024role, ward2026correcting, creed2026quantum} often use relatively recent high quality crystal structures\cite{xiao2020single} (CSD code OHUBUR). 
Here we use dimers extracted from this crystal structure, following the approach and numbering scheme of Giannini~et al.\cite{giannini2024role}. 
Dimers D1--D6 all have a minimum intermolecular distance of less than $3.5 \AA$. 

\begin{figure*}[htbp]
    \centering
    \includegraphics[width=\textwidth]{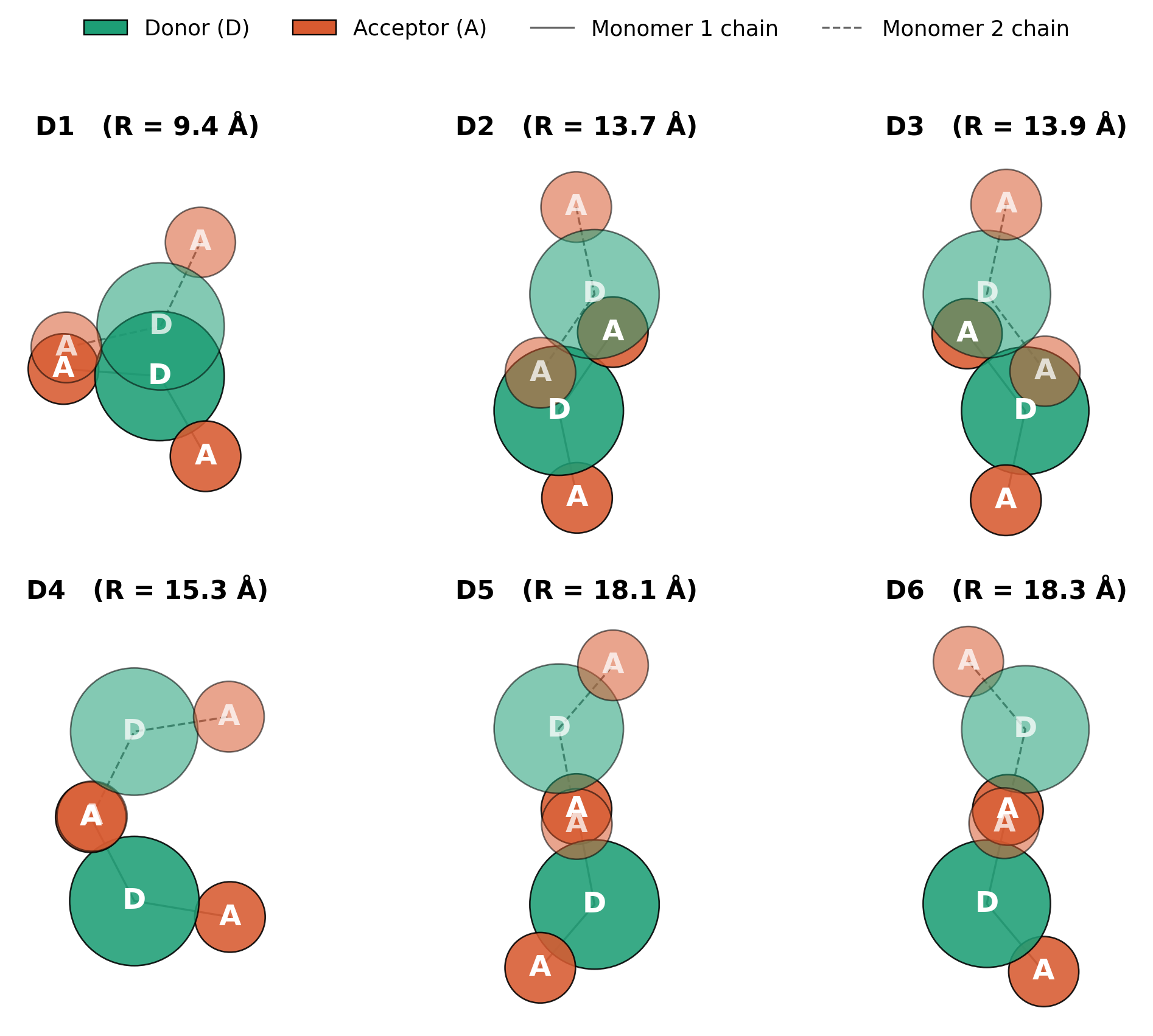}
    \caption{Overlapping nature of the donor and acceptor moieties in the six dimer geometries.}
    \label{fig:dimer-ada}
\end{figure*}

We found that the nonadiabatic dynamics followed the nature of the dimer occlusion: `DA dimers' D2 and D3 which have multiple overlapping donor and acceptor units between the two monomers; `AA dimers' D4--D6 whose overlap is via a single acceptor-acceptor moiety;  and D1 where both one donor fragment and one acceptor fragment on the monomer overlap with the same fragments on the other monomer.
The overlapping nature of these dimers is shown in figure \ref{fig:dimer-ada}. 

% needs fixing and removing the subfigure stuff

% Jarv to redraw this figure on top of the actual dimer structures

Recently\cite{ward2026correcting} we characterised and ordered the charge transfer (CT) and Frenkel Exciton (FE) singlet and triplet states in the Y6 dimers. 
Most photoexcitation goes into a bright (large oscillator strength) state, which in organic semiconductors is generally of Frenkel Exciton character\cite{PopeSwenberg1999}.
We\cite{creed2026quantum} constructed a 5-state effective model (parametrised by a dimer projection method and explicit calculation of inter-system crossing matrix element) and then simulated non-adiabatic dynamics using the Hierarchical Equations of Motion (HEOM), starting with the singlet FE state populated. 
We show that triplets in this model are mainly generated via a transient \textit{intermolecular} charge-transfer singlet to triplet Frenkel exciton route on a time-scale of $> 1 $~ps, yet transfer to the CT state within the singlet manifold occurred rapidly ($\approx 100$~fs). 

An effective model loses much chemical specificity, both through reducing (averaging) the manifold of electronically excited states (i.e. not fully describing the FE-CT mixing we calculated in our previous study\cite{ward2026correcting}), and simplifying vibronic coupling to a harmonic bath with an associated spectral function.

Here we therefore directly simulate the atomistic nonadiabatic dynamics of these dimers, to gain understanding of the short-time ($< 100$ fs) internal conversion on the singlet manifold, including anharmonic and collective responses of the nuclear degree of freedom. 
We neglect both inter-system crossing to the triplet manifold and decay to the ground state (i.e. fluorescence), as these processes do not occur on this time scale.

We simulate the nonadiabatic dynamics with fewest switches surface hopping (FSSH), where the nuclear degree of freedom is updated with a $0.5$~fs integration timestep. 
Due to the large size of the dimers (170 atoms, without the full size sidechains), for computational efficiency we use hybrid time-dependent DFTB (TD-DFTB)\cite{Hourahine2025, DazMirn2024}, with the recently developed OPV parameters\cite{sun2025hybrid}. 
Unlike previous atomistic surface-hopping simulations of crystalline Y6 which used a parametrised Frenkel-exciton manifold\cite{Giannini2022Exciton}, our on-the-fly hybrid TD-DFTB approach explicitly resolves coupled intra- and intermolecular charge-transfer states and calculates explicit analytic nonadiabatic coupling vectors\cite{niehaus2021ground,niehaus2023exact}.

We find that all six Y6 crystal dimers relax rapidly toward $S_1$ within \qty{100}{\femto\second}, with electron and hole moving onto separate molecular fragments (a charge transfer state). 
Packing geometry decides whether that separation is useful from a device physics point of view. 
In donor--acceptor dimers ($D_2, D_3$), overlap drives the exciton into an intermolecular CT state, even as the electron--hole distance reduces; in acceptor--acceptor dimers ($D_4$--$D_6$), electrostatic repulsion across the sole A--A contact drives the excitation into mixed intra- and intermolecular states, with the face-on dimer ($D_1$) straddling the two limits.
Capturing this dynamic competition required explicit atomistic electronic structure and nuclear degrees of freedom. We suggest that this effect is directly linked to the observed polymorph-dependent charge generation in Y6\cite{Kroh2022-wn}.

\section{Methods}

\subsection{Electronic structure method} 

 Due to the large size of the dimers (170 atoms, without the full size sidechains), for computational efficiency we use Density Functional Tight Binding (DFTB)\cite{Elstner1998}, in combination with its hybrid functional formalism\cite{van2023hybrid} and its linear-response time-dependent extension (TD-DFTB)~\cite{Niehaus2001,kranz2017time} as implemented in the DFTB+  code\cite{dftbpluscode}, with the recently developed B3LYP OPV parameters\cite{sun2025hybrid} in our FSSH dynamics and use the DFTB3 extension\cite{Gaus2011} in our ground state  molecular dynamics (GS MD) simulations which generate the initial conditions for the dynamics.

We validate the use of this method in describing the electronic states of the Y6 dimers by comparing the properties of the excited states generated by the method to those obtained by TDDFT calculations with our optimally tuned\cite{ward2026correcting} range-separated hybrid functional (OT-RSH) which we believe well describes the excited states of these Y6 dimers. 
We show this analysis in the supplementary information (Figures S1-S4).  

\subsection{FSSH dynamics} 

We perform nonadiabatic molecular dynamics (NAMD) simulations \cite{Jain2022, Subotnik2016, Mukherjee2024, Wang2016} within the fewest switches surface hopping framework \cite{tully1990molecular}, using the recent TD-DFTB implementation \cite{diaz2024non}. Tight-binding methods have been used extensively for NAMD simulations across a wide variety of systems and provide a reliable description of their photochemistry \cite{diaz2024non,pal2016nonadiabatic,stojanovic2017nonadiabatic,diaz2024exploring,akimov2016nonadiabatic}. All simulations were carried out with DFTB+\cite{dftbpluscode} interfaced with SHARC3\cite{mai2018sharc,mai2023sharc3}.

Initial conditions for each dimer were generated by running  a set of ground state molecular dynamics trajectories at 300K, each trajectory was run for  1 ps with a 0.5 fs timestep. This interval was chosen as it is well beyond the decorrelation time of the nuclear velocities (Figure S5 in the SI ). During the ground state dynamics, selected atoms were kept fixed so that the inter-monomer distances in each dimer remained consistent with the crystal structure, these constraints were then relaxed when later running the NAMD trajectories. The precise atoms used to fix the intermolecular distance for each dimer are shown in table S1 in the supplementary information. 

For each initial condition, we computed the first four singlet excited states ($S_1$--$S_4$) with TD-DFTB. Each trajectory was then started in the brightest state, identified as the one with the largest oscillator strength, which corresponds to the locally excited (LE) state.

NAMD trajectories were then propagated for \qty{100}{\femto\second} with a 0.5 fs nuclear timestep and 25 interpolated electronic substeps. 
Hopping probabilities were computed from nonadiabatic coupling vectors between all pairs of states, evaluated analytically \cite{niehaus2021ground,niehaus2023exact}. 
%\jarv{Should probably flag this up in the intro, where introducing TD-DFTB... most semiconductor work is with the LZ approx.}
Decoherence was accounted for with the energy-based decoherence correction \cite{granucci2007critical}, using a decoherence parameter of 0.1 Hartree. A total of 50 trajectories per dimer were run.

\subsection{Theodore analysis} 
\label{subsec:theodore}
Throughout this paper, we use TheoDORE\cite{plasser2020theodore} to analyse the properties of the current adiabatic state of the fewest switches surface hopping (FSSH) trajectories. TheoDORE calculates state properties by considering the ratio of the diagonal and off-diagonal elements of, $D_{ab}^{0i}$, the ground to $i$-th excited state transition density matrix (TDM) in the orbital basis $\{a,b\}$. The orbitals $\{a,b\}$ are localised on the molecular fragments $A$ and $B$ \cite{plasser2020theodore}. For the Y6 dimers, one natural motif for the fragments is the two monomers. 
In addition, we further subdivide A-DA'D-A Y6 into three fragments: acceptor end units (A) and a donor core (D, encompasing the central fused DA'D core). 
This now ADA structure is shown in figure \ref{fig:dimer-ada}, fragmenting the dimer into six separate units. 

One key property we want to predict is the charge-transfer (CT) character.  In TheoDORE\cite{plasser2012analysis,plasser2020theodore}, the CT character corresponds to the proportion of excitation that occurs between fragments, the degree to which the excitation leads to charge transfer. Additionally, Theodore \cite{plasser2020theodore} can calculate using the TDM the average root-mean squared distance of electron-hole  across the different atoms in a given excited state, $RMSeh$. 

For our dimers, an important charge transfer character  to examine to enable us to determine whether the current state of the trajectory is a Frenkel Exciton (FE) or a Charge Transfer (CT) state is that obtained by separating the dimer into its two monomer units.  We will, here, refer to this CT character as the intermolecular CT character, $\omega_{CT}$ (or $\omega_{CT}^{Int}$ in the discussion below).   A value of $\omega_{CT}=1$ therefore corresponds to a `pure' CT state, while $\omega_{CT} = 0$ is a `pure' FE state.  Due to CT-FE state mixing at short separations most excited states of the Y6 dimers are not pure FE or CT states.  Following the works by Zheng~et al.\cite{zheng2017effect} and Ward~et al.\cite{ward2026correcting}, we consider CT states as those states with $\omega_{CT} \gtrsim 0.75$ and FE states as those states with $\omega_{CT} \lesssim 0.25$ with all other states considered mixed states. 

To gain further insight into the excited states of the current adiabatic state of the trajectory, we can also examine the CT character by separating the dimer into six rather than two fragments and then calculating the $\omega_{CT}$ value. This CT character gives the charge-separation both across different (ADA) fragments of a given monomer unit and between monomer units. We refer to this as the Intramolecular+Intermolecular (Intra+Int) CT character, $\omega_{CT}^{Int+Intra}$. Given that $\omega_{CT}^{Int+Intra}$ is calculated by adding the off-diagonal matrix elements, it can clearly be broken down into two separate entities; the intermolecular CT character ($\omega_{CT}^{Int}$ discussed above) and an intramolecular CT character ($\omega_{CT}^{Intra}$) that measures the extent to which a state is charge-separated within the individual monomer units (ADA), where 
$\omega_{CT}^{Intra+Int}=\omega_{CT}^{Intra} +\omega_{CT}^{Int}$.

\section{Results}

\subsection{Population Dynamics}

\begin{figure*}
    \centering
    \includegraphics[width=1\linewidth]{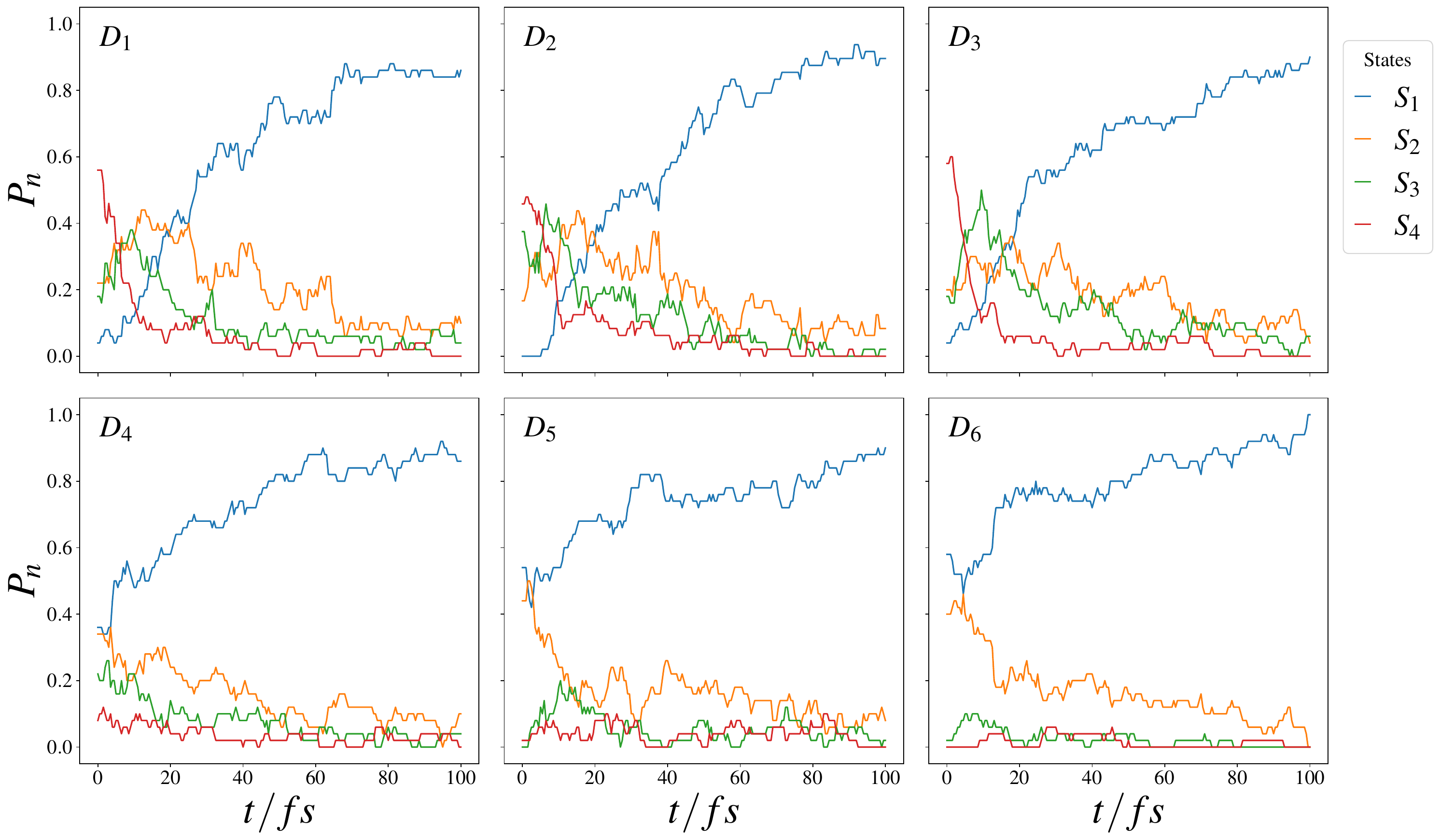}
    \caption{Time evolution of the population of the different excited states (indicated in plot) averaged over 50 trajectories for the different contact pair dimers of Y6. Top panel left to right D1-D3. Bottom panel left to right D4-D6.  }
    \label{Population_dynamics}
\end{figure*}

 For a given dimer and set of initial conditions in our FSSH simulations we photoexcite into the brightest (largest oscillator strength) state, which in organic semiconductors is generally of Frenkel Exciton character. From the FSSH dynamics, we can examine how the population of the different singlet states ($S_1$--$S_4$) evolves with time for the different dimers ($D_1$--$D_6$) shown in  figure \ref{Population_dynamics}. This figure shows that in the donor--acceptor dimers ($D_2, D_3$) and the face-on dimer ($D_1$), exciting into the optically bright FE state results in population transfers from predominantly the S3 and S4 states at $t=0$~fs to mainly the S1 state at  $t=100$~fs  after photoexcitation. By contrast,  for the acceptor--acceptor dimers ($D_4$--$D_6$), this excitation into the most optically bright FE state results in population transfer from predominantly the S2 and S1 states at $t=0$~fs to primarily the S1 state by 100~fs after excitation. 
 This difference in initial state reflects the relative adiabatic state ordering of the dimers: in DA and face-on geometries, optically dark intermolecular CT states lie below the bright FE manifold, pushing photoexcitation up into $S_3$ and $S_4$. 
 In our dynamics no population transfer occurs to the S0 state due to the large bandgap.

\subsection{Understanding time evolution of CT character}

\begin{figure*}
    \centering
    \includegraphics[width=1\linewidth]{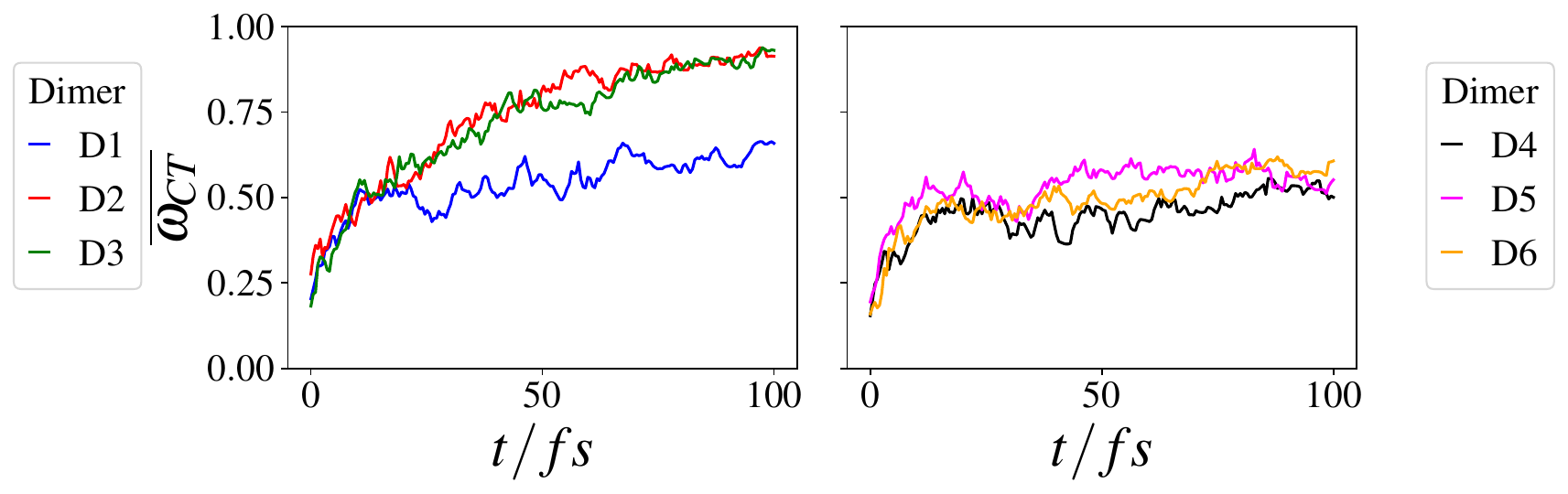}
    \caption{Time evolution of the average intermolecular charge-transfer character ($\bar{\omega}_{CT}$)  of the current state of the trajectory averaged over 50 trajectories for the different Y6 contact pair dimers (indicated in plot).}
    \label{Charge_dynamics}
\end{figure*}

In order to further understand this population transfer we examine in Figure \ref{Charge_dynamics}
how the intermolecular-charge transfer character (obtained by fragmenting the dimer into 2 fragments and discussed in section \ref{subsec:theodore}) of the current adiabatic state of the trajectory averaged over trajectories varies with time for the different contact pair dimers of Y6.  We find that for the donor--acceptor dimers ($D_2, D_3$) there is clear evolution of the system from a FE state ($\overline{\omega_{CT}}<0.25$) to a CT state ($\overline{\omega_{CT}}>0.75$) during the trajectories. By contrast, for the acceptor--acceptor dimers ($D_4$--$D_6$) we see that  
the CT character of the states only slightly increases during the dynamics from an FE to mixed FE-CT state.  The face-on dimer ($D_1$) straddles these two cases, where it shows relatively clear evolution from an FE to a CT state  throughout the dynamics, but has far greater FE-CT hybridisation than the donor--acceptor dimers.

\begin{figure*}
    \centering
    \includegraphics[width=1\linewidth]{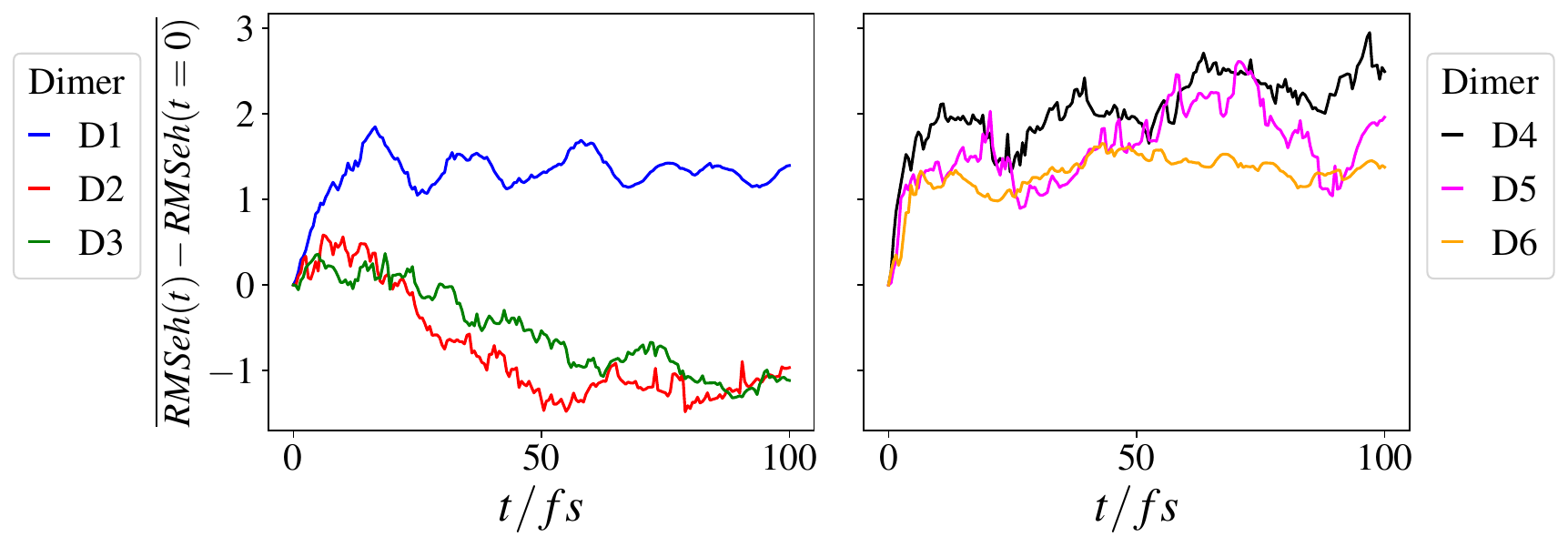}
    \caption{Time evolution of the average root mean squared distance between the electron and the hole of the current state of the trajectory relative to it value after the initial excitation, averaged over 50 trajectories for the different Y6 contact pair dimers (indicated in plot).}
    \label{root-mean-square}
\end{figure*}

Another natural quantity to examine is how the trajectory-averaged electron-hole root-mean-square distance, $RMSeh$, changes with time (see figure \ref{root-mean-square}). 
We find that this does not correlate well with the intermolecular CT character, the key parameter from a device physics perspective to facilitating the formation of free charges at later times. 
For example, in the donor--acceptor dimers ($D_2, D_3$), trajectories show a clear decrease in the e--h distance during FE to CT evolution, while in the acceptor--acceptor dimers ($D_4$--$D_6$) the e--h distance increases with only a small increase in CT character.  
The root mean squared distance, $RMSeh$ clearly reflects the geometries of these dimers and in particular their stacking rather than being a predictive quantity to determine whether free charges can be formed from these dimers in devices, which has previously been observed for a different system both experimentally and theoretically by M\"uller~et al.\cite{Muller2025}.

\begin{figure*}
    \centering
    \includegraphics[width=1\linewidth]{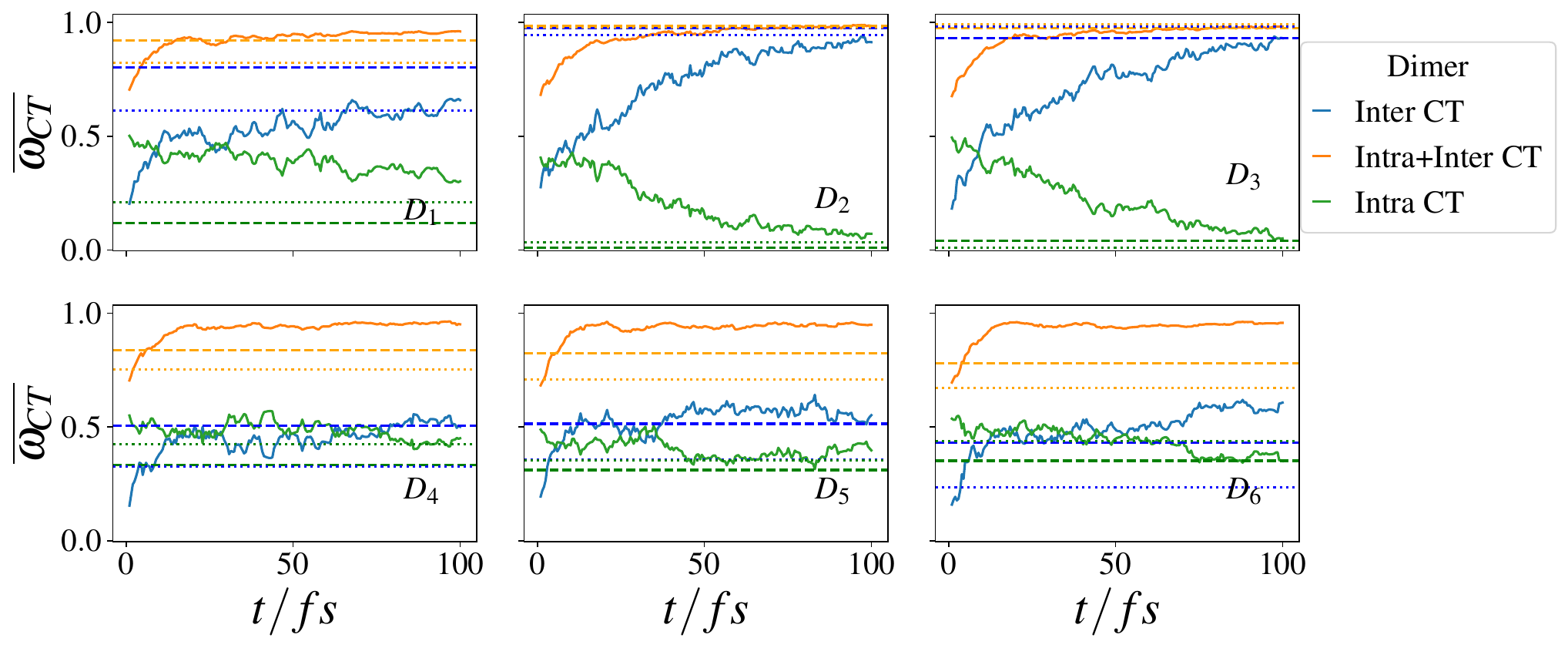}
    \caption{Time evolution of the intermolecular CT character (charge separating across the two different monomer units), intra- and inter- molecular CT character (charge separating into any of the 6 ADA' fragments), and intra molecular CT character (charge separating between different fragments on the same monomer) for the current state of the trajectory averaged over 50 trajectories for the 6 contact pair Y6 dimers. Top panel left to right D1-D3. Bottom panel left to right D4-D6. The dashed lines in the figures show the average intra/intra+inter/inter molecular CT character of the S1 state of the different initial structures of the dimers at $t=0$ (sampled by GS MD) and the dotted lines show the intra/intra+inter/inter molecular  CT character of the S1 state of the original crystal structure dimers obtained by Giannini~et al.\cite{giannini2024role}. }
    \label{Charge_dynamics_int_intra}
\end{figure*}

% Jarv - I changed the above description to reflect what is written in the caption (3rd one being intra), but I think perhaps what you calculated was the third one is the diagonal excitation, i.e. Frenkel exciton like?

To further understand the charge-separation dynamics in the Y6 dimers, we can calculate the three different types of CT character obtained by breaking the Y6 dimer into 6 fragments (discussed in section \ref{subsec:theodore}). These different CT characters are the CT character across all 6 fragments (intra+intermolecular CT character), CT character between the two monomer units (intermolecular CT character) and the charge separation across the individual monomer units (intra molecular CT character). The time evolution of all three CT characters of the current state of the trajectory averaged over trajectories for the different dimers ($D1$--$D6$) is shown in figure \ref{Charge_dynamics_int_intra}. 
We demonstrate that our DFTB method correctly calculates the strength of electron-donating and electron-accepting moieties in Y6 by comparison to constrained DFT\cite{nwchem} (supplementary information Tables S2 and S3). 

Figure \ref{Charge_dynamics_int_intra} shows that for the donor--acceptor dimers ($D_2, D_3$)  the system initially has some intra-molecular CT character in the FE state obtained after photoexcitation, but by 100~fs all charge-separation occurs as intermolecular CT character across the two monomer units, as these dimers go from clear FE to CT states in the dynamics. This is facilitated by the structures of these dimers, which have DA overlap allowing the favourable splitting of the electron and hole onto the A and D units respectively.  By contrast, figure \ref{Charge_dynamics_int_intra}  shows that for the acceptor--acceptor dimers ($D4$-$D6$)  while the intermolecular CT character increases during the dynamics, the system retains some intra-molecular CT character at the end of our simulation window (100 fs) as they form  mixed CT-FE states.  This occurs because there is competition between forming a CT state across the D-A unit of the monomer where the hole and electron can sit on their preferred moiety and the A-A overlap in the dimer where the hole has to unfavourably sit on an acceptor unit to remain relatively close to the electron when charge-separates across the monomer units.  We again see that the face overlapping dimer (D1) straddle the two other cases. This occurs because  while there is the greatest degree of overlap in this dimer out of all the dimers $D1$-$D6$, this overlap is  via A-A  and D-D overlap which are both less favourable for charge separation.

Figure \ref{Charge_dynamics_int_intra}, however, shows that for all dimers by 100 fs the electron and hole are on separate fragments, just depending on the type of overlap in the dimers determines how much of this charge separation occurs within the same monomer unit or across monomer units.  From figure \ref{Charge_dynamics_int_intra} it is clear that in order to get the largest degree of intermolecular charge separation in Y6 the D-A overlap is required. This effect of the dimer structures on the extent of intermolecular CT generation gives some insight into the polymorph-dependent charge generation in Y6\cite{Kroh2022-wn}.

Figure \ref{Charge_dynamics_int_intra} also examines the  extent to which the average intermolecular, intramolecular, and intermolecular+intramolecular CT character of the ensemble of the trajectories at 100 fs is captured solely by the CT character of the S1 state of the initial dimers of Giannini~et al.\cite{giannini2024role} (dotted lines in plot) or the average CT character of  the S1 states of the structures sampled with GS MD that are the input structures of our FSSH trajectories at $t=0$ (dashed line in plot). We here find that the CT characters of  S1 state of both the crystal structure dimers and the initial structures used as input to our FSSH trajectories clearly play an important role in determining the properties of the various Y6 dimers at $t=100fs$.  For example, both  predict the distinctly different behaviour of the different types of Y6 dimer. Both quantities, however, do not fully describe the properties of the dimers at $t=100fs$. This occurs for two reasons. Firstly, from figure \ref{Population_dynamics} there is population of not just the $S_1$ state but also higher excited states, e.g. the $S_2$ state.  Additionally,  in some trajectories in the dynamics there are state crossings which change the order and properties of e.g. the $S_1$ and $S_2$ states as the dimers geometries subtly  change during the dynamics. In the Y6 dimers these affects are particularly relevant as there are small energy gaps between the FE/CT/mixed FE-CT states which mean very small changes in geometry change the ordering.  As a result, understanding the dimers properties at $t=100fs$ after photoexcitation is not as simple as understanding the properties of the stationary structures  $S_1$ state's.

Generally speaking, the GS MD sampled S1 state structures average CT characters are a better reflection of the average CT characters obtained from our trajectories at 100fs than the CT characters of the S1 state of the original crystal dimers \cite{giannini2024role}. This is not that surprising given that Wigner-sampled ground-state $S_{0}$ structures are routinely used as a cheap, zero-order approximation for the $S_{1}$ emission spectrum \cite{CrespoOtero2012, Fehr2021}. It is however, known that this approach can fail to predict emission spectrum accurately due to  missing the change in geometry in the excited state \cite{Barbatti2015, McCoy2021,  PinheiroJr2023} and due to the effects of state crossings during the dynamics\cite{Sre2020, Kossoski2018}. A key take home message from this work, is that a full understanding of the charge transfer character of the Y6 dimer excited states following photoexcitation can only be achieved through running non-adiabatic dynamics and not simply from examining the stationary structures. 

 \begin{figure}
    \centering
    \includegraphics[width=1\linewidth]{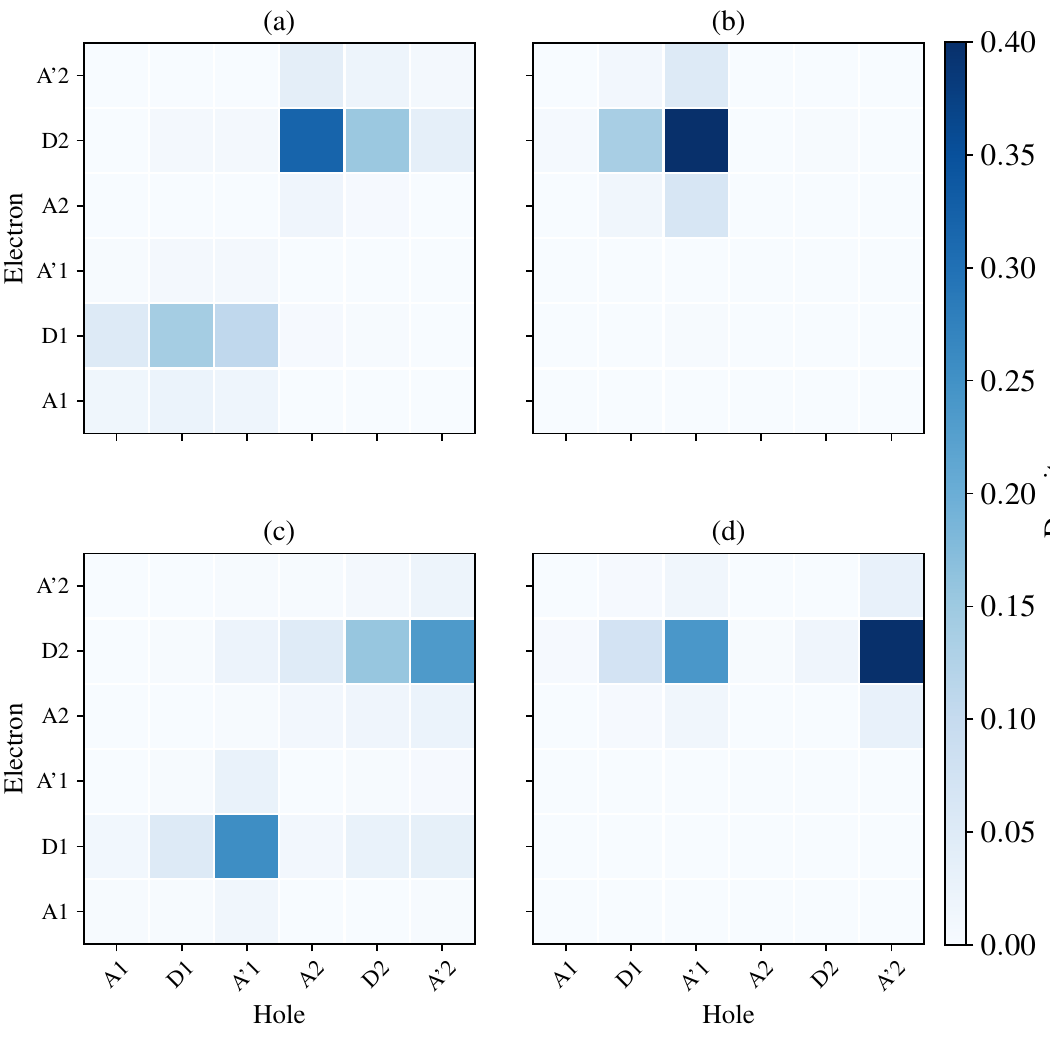}
    \caption{Electron/Hole correlation plots of the site an example D2 dimer trajectory which was excited into $S_4$ state at $t=0fs$ (a) and the final state of the trajectory at $t=100fs$ $S_2$ (b),  an example D4 dimer trajectory which was excited initially into the $S_2$ state at $t=0fs$ (c) and the final state of the trajectory at $t=100fs$ $S_1$ (d) . Monomers within each dimer are fragmented into two (acceptor) end units and a core (donor) unit, denoted as $A/A'$ and $D$ respectively. The colour bar again corresponds to the 1 particle transition density (given in terms of the electron and hole position) for the given excited state. Plots of the natural orbitals of these states can be found in the supplementary information.}
    \label{D2_D4_ecample_traj}
\end{figure}
 
 To further exemplify the difference in dynamics of the different types of dimer, figure \ref{D2_D4_ecample_traj} shows the e-h plots for the initial and final state of a typical donor--acceptor dimer ($D2$)  and acceptor--acceptor dimer ($D4$) trajectory. This clearly shows that the for a donor--acceptor dimer's trajectory there is clear evolution from an FE to CT state during the dynamics while for the acceptor--acceptor dimer's trajectory the system starts in FE like state and goes to a mixed FE--CT state during the dynamics.  In the supplementary information (Figures S6--S9) NTO electron-hole plots are shown for both dimers initial and final states to further show this. 

In our previous work \cite{creed2026quantum} we examined the the transfer of population between different states in the Y6 dimers using a simple 5-state effective model. Here we defined an effective FE state and CT state and examined the transfer rate between them using electron and hole coupling matrix elements calculated using the counterpoise method\cite{Kirkpatrick2007, Baumeier2010}, and found that the transfer between these states occurs on a time scale on the order of ${\cal{O}}(\qty{100}{\femto\second})$. In this work, while we see that the dynamics between the various singlet states in Y6 occurs on the time scale predicted by the effective model, the actual dynamics is a far richer picture than simply transfer between the FE and CT states. In reality, the chemical structure and in particular the orientation of the ADA fragments of the dimers plays a strong influence on the nature of the dynamics and the types of state formed. For the donor--acceptor dimers and face on dimer, their dynamics can essentially be thought of as intermolecular FE to CT transfer which is at least qualitatively captured by an effective model. For the acceptor-acceptor dimers, the type of charge-separation and the time evolution of the states formed after the initial excitation is far more subtle and can only be fully captured by a fully atomistic model.

This work also provides microscopic insight into the polymorph analysis of Kroh~et al.\cite{Kroh2022-wn}. 
Mapping their core--terminal packing classification onto our contact pairs, our donor--acceptor dimers ($D_2, D_3$) correspond to their ``core--terminal'' pairs, the face-on dimer ($D_1$) to their ``core--core / terminal--terminal'' pair, and our acceptor--acceptor dimers ($D_4$--$D_6$) to their ``terminal--terminal'' pairs.
%(In the rest of this paragraph, following Kroh et al., 'C' denotes 'core' and 'T' denotes 'terminal'.) - now redundant as I rewrote it all (?) hopefully correctly :^)
% Jarv - above was very confusing, I tried to edit it to make it better
Kroh~et al.\cite{Kroh2022-wn} found that morphologies rich in donor--acceptor pairs generate free charges more efficiently than face--on pairs, which in turn outperform acceptor--acceptor. 
We provide a starting point to understanding this, by noting that these  dimers have differing fast generation of intermolecular charge transfer, which we propose then leads to charge separation at later times.  
The competition between forming intramolecular and intermolecular charge separation due to the difference in the arrangement of the fragments cannot be understood from a tight binding Hamiltonians constructed from the monomer's FE and CT states,  but only occurs through considering the atomistic structure of these dimers. 
% Jarv - really need to fix the talk above, super confusing, cannot accept having 'CT' mean anything other than charge-transfer in this paper really. 
% OK, quickly tried my best, still a little messy

As part of our work, we also have conducted normal mode analysis with SHARC 3\cite{mai2018sharc, mai2023sharc3}, see supplementary information. 
We find that there is significant excitation of the low frequency modes that move the two monomer units of the dimer relative to each other (Peierls distortions). 
We also identify distinct intramolecular vibronic bands at $750cm^{-1}$ and $1000$--$1500 cm^{-1}$.  
We do not find any clear differences between the dimers. 
Resolving dimer-specific vibrational variation, and getting sufficient statistics to identify and understand the slow intermolecular modes, will require significantly larger ensembles than we were able to access within our computational budget. 

\section{Conclusion and Future work}

In this work, we have studied the short-time dynamics of the contact pair dimers of Y6 after photoexcitation. We found that the nature of this dynamics  followed the nature of the dimer occlusion; separating dimers into three distinct categories donor-acceptor dimers,  face on  dimers, and acceptor--acceptor dimers.  We found that the donor--acceptor dimers are better than the face on dimers which are better than the acceptor--acceptor dimers on building up intermolecular charge transfer character, which is integral to forming free charges in real devices.  This difference in behaviour of the different dimers helps explain the polymorph dependence of charge generation in Y6. 

Additionally, we found that in order to understand the charge transfer character of the Y6 dimers 100fs after photoexcitation full non-adiabatic simulations must be run, as this is not correctly predicted by simply looking at the stationary properties of $S_1$ states of either the crystal or ground-state molecular dynamic structures. 

If we were able to run dynamics faster, there are many different directions that we could extend our study. 
For reference, a single 100~fs of dynamics takes around 100 wall-clock hours on a 64-core Intel Xeon 8358 node (Imperial CX3 HPC cluster\cite{HPC}), with most of this time in calculating the non-adiabatic coupling matrix elements. 

With larger aggregates we could more closely model the solid state, and undertake simulations with explicitly applied fields to look at charge separation. 
With better statistics we could more completely understand the nature of the vibrational modes and how they drive the observed difference in dimers. 
We could include the full side-chains, which may modulate the dynamics. 
Larger length and time scales would produce the data for developing tight-binding models (following \cite{Giannini2022Exciton}) to model very large aggregates.

%\charly{In my opinion, we need to add more chemical prospective like outlooks in the conclusions.}
% Jarv - I think this is a very good idea! At least at the level of understand what chemical probes may change the behaviour ,and therefore provide some spectrscopic comparison etc.

\begin{acknowledgements}
% thank PEOPLE first, then funding
The authors would like to thank Jenny Nelson, Tim Rein, Thomas Ward, Jolanda M\"uller, Thomas Niehaus and B\'{a}lint Aradi for their helpful discussions.  
I.C. is supported by EPSRC (EP/Y020790/1)
J.M.F. is supported by a Royal Society University Research Fellowship (URF-R1-191292).
E.W. is a Royal Society funded PhD student on the above grant (URF-R1-191292). 
We gratefully acknowledge the use of the Imperial College Research Computing Service\cite{HPC}.
Additionally, via our membership of the UK's HEC Materials Chemistry Consortium (MCC), funded by EPSRC (EP/R029431 and EP/X035859), this work used the \textsc{ARCHER2} UK National Supercomputing Service (http://www.archer2.ac.uk).
\end{acknowledgements}

\bibliography{ref.bib}

\end{document}